\documentclass[doublespacing]{elsart}

\usepackage{graphicx}

\usepackage{amssymb}

\usepackage{lineno}
\usepackage{color}

\begin{document}

\begin{frontmatter}



\title{Adsorption and onset of lubrication by a double-chained cationic surfactant on silica surfaces}


\author{Laurence Serreau, Muriel Beauvais, Caroline Heitz, and Etienne Barthel  }
\ead{etienne.barthel@saint-gobain.com}
\address{Laboratoire Surface du Verre et Interfaces, UMR 125, CNRS/Saint-Gobain, 39 quai Lucien Lefranc, B.P. 135, F-93303 Aubervilliers cedex, France}

\begin{abstract}
In the context of glass fiber manufacturing the onset of lubrication
by a C$_{18}$ double-chained cationic surfactant has been
investigated at high normal contact pressures. Comparison with
adsorption kinetics demonstrates that lubrication is not directly
connected to the surfactant surface excess but originates from the
transition to a defect-free bilayer which generates limited
dissipation. The impact of ionic strength and shear rate has also
been studied.
\end{abstract}

\end{frontmatter}
\linenumbers
\section{INTRODUCTION}

During glass fiber manufacturing, the high friction characteristic
of silicate surfaces in water results in surface damage and
eventually prejudices the tensile strength of the fibers. The
necessary lubrication can be imparted through an aqueous dispersion
(sizing) which is applied at the initial stage of the glass fiber
manufacturing process. The sizing serves many purposes but
double-chained cationic surfactants (softeners) are often added to
this dispersion to participate in lubrication. However, it is well
known that adsorption of surfactants proceeds
slowly~\cite{Chen92,Biswas98}, especially for long-chained
amphiphiles~\cite{Hayes98}. The question we address in this paper is
the kinetics of lubrication: once surfactant adsorption has started,
when will lubrication be effective ?

It is expected that the answer depends upon the mechanical loading
and the friction velocity in a complex manner. In practice, the
typical drawing speed is several meters per second, but the contacts
between the several hundred glass filaments within one fiber will
slide at much slower velocities, which can be in the range of
millimeters per second or lower. \textcolor{black}{The filaments slide
against each other in the presence of the sizing which initiates
both adsorption and lubrication. Such are the operating conditions
we emulate in the present study}.

Numerous studies have been conducted on the contact and also the
friction properties of surfactant covered surfaces. For practical
reasons, the bulk of the literature is devoted to short, single
chain surfactants, which exhibit faster
equilibration~\cite{Subramanian01,Vakarelski04,Oncins05}.
Simultaneously, because of their relevance in biological
applications, numerous papers deal with the adsorption of lipids. In
particular the structure of the surface aggregates and the
mechanical response of these insoluble double-chained surfactants
have been studied in great
details~\cite{Oncins05,Pashley86,Grant02,Boschkova02,Boschkova02b}.

\textcolor{black}{In this paper, we investigate the early stages of
lubrication just after immersion of silica surfaces in an aqueous
dispersion of a typical double-chained (2~C$_{18}$) cationic
surfactant. The surfactant dispersion was investigated by Small
Angle Xray Scattering (SAXS) and Static Light Scattering (SLS) and
the adsorption kinetics on silica surfaces by Attenuated Total
Reflection (ATR) Fourier Transform Infrared (FTIR) spectroscopy.
Using atomic force microscopy (AFM) and macroscopic friction tests,
we have measured the contact properties (repulsive barrier, adhesion
and friction) of macroscopic silica surfaces in the initial stages
of adsorption as a function of time after immersion. The results
highlight the impact of the adsorption kinetics
\textcolor{black}{and the changes of the surfactant configuration during the early stages of adsorption.
The picture which emerges is that of a gradual transition from a disordered adsorbed
layer with high friction to a lubricating defect-free bilayer.
Shear is shown to play a role in the transition to the
lubricating state.}}
\section{EXPERIMENTAL SECTION}
\subsection{Materials}
\begin{figure}\begin{center}
\includegraphics[width=4cm]{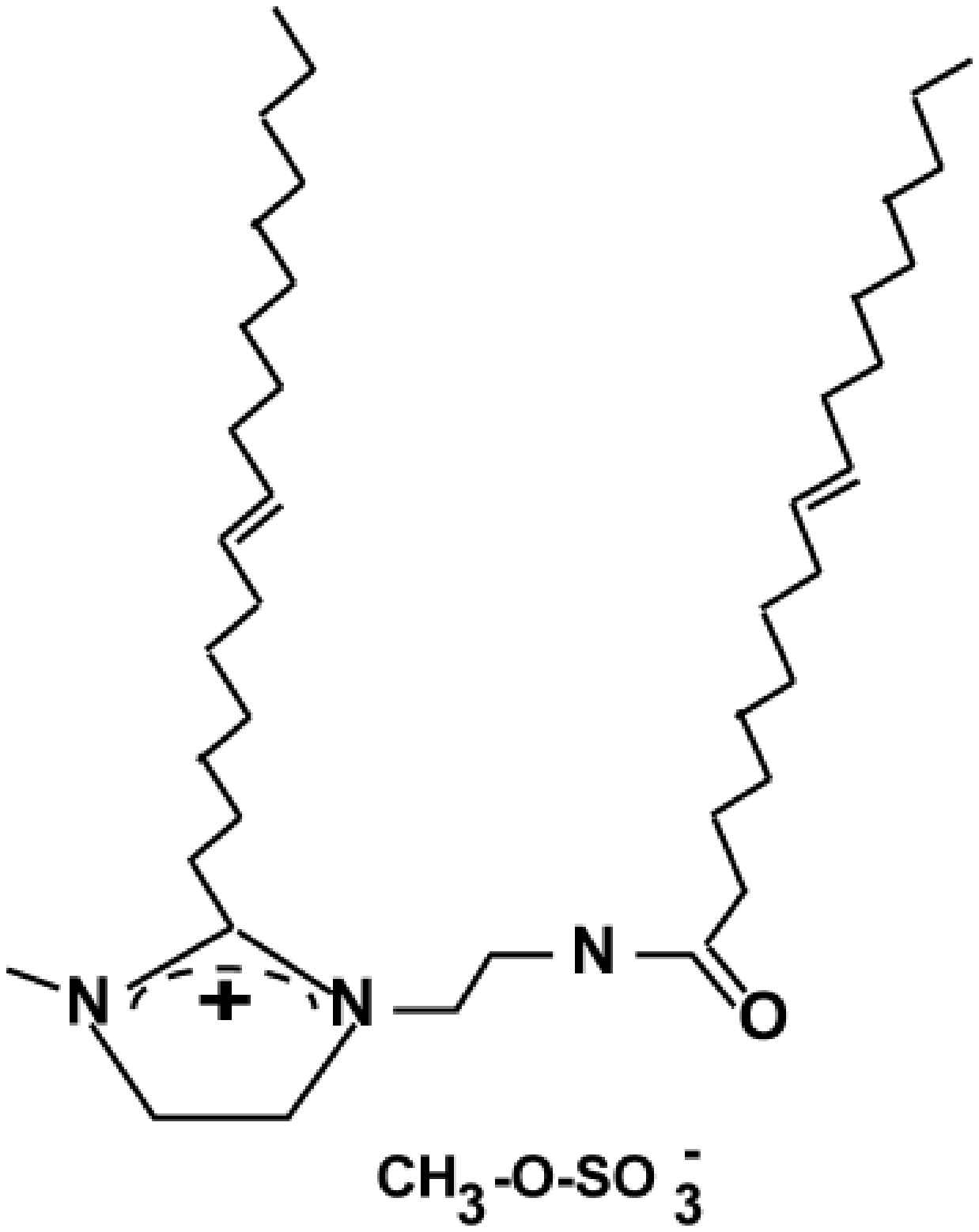}
\caption{Chemical structure of the DOAIM}\label{Fig_DOAIM}
\end{center}\end{figure}

The double-chained cationic surfactant
1-methyl-2-noroleyl-3-oleic acid-aminoethyl-imidazolinium
methosulfate (DOAIM, Figure~\ref{Fig_DOAIM}) in isopropanol (25\%
wt) is obtained from Goldschmidt Rewo GmbH \& Co.,(Germany) and used
as received. The molecular weight is 740 g/mol and the density
0.97~g/cm$^3$. The CMC \textcolor{black}{with isopropanol} measured by surface tension is 1$\times$10$^{-5}$~M.
The chain melting temperature is 46$^\circ$C as measured by DSC, in
agreement with the values obtained for similar
compounds~\cite{Liu96}. All the experiments were performed at
ambient temperature.

Solutions of DOAIM at 5$\times$10$^{-4}$~M or 1$\times$10$^{-3}$~M were prepared
in milli Q water with 24 hours gentle stirring after evaporation of
the isopropanol at 60°C. Most experiments were carried out at
natural pH$\simeq 4.6$. In a set of experiments, the ionic strength
was varied with acetic acid/sodium acetate while maintaining
constant pH=4.5. Such concentration and pH conditions are typical
for actual sizing formulations.
%
%
\subsection{Methods}
\subsubsection{Equilibrium characterization}
\textcolor{black}{The SAXS experiments were performed in a Kratky
set-up (Anton Paar) with a Cu K$_\alpha$ source (0.1542~nm) and a
linear gaz detector placed at 23~cm from the source. The SLS
experiments were performed on a Malvern Zetasizer 3 equipped with an
He-Ne laser (633 nm), a photomultiplier and a goniometer. The same
piece of equipment was used to measure the zeta potential by
electrophoretic mobility in a liquid cell. The laser interferometric
comb method was used. The test system was 200 nm diameter silica
particles (St\oe ber synthesis). Adsorption at a given concentration
was carried out by dilution from a 10$^{-4}$~M surfactant solution
followed by 5 hour equilibration time. Surface tension was measured
by the Wilhelmy plate method.}

\subsubsection{FTIR/ATR adsorption kinetics}
Adsorption kinetics were measured by FTIR spectroscopy in the ATR
mode using a Nicolet Nexus 670 spectrometer equipped with a MIR
source, a KBr beamsplitter and a MCT-A detector. The experiments
were carried out on a germaniun internal reflection element
(trapezoidal, 50$\times$10$\times$1mm$^3$, 45$^\circ$ incident
angle) covered on the larger side by a silica layer $\simeq$7 nm
thick deposited by magnetron sputtering. Before use, the surfaces
were cleaned with a sequence of detergent solution, deionized water,
acetone and absolute ethanol for 15 minutes in an ultrasonic bath,
followed by a final UV/Ozone treatment for 1 h. After cleaning, the
wafer was introduced in the internal multi-reflection cell which was
immediately assembled and aligned in the sample compartment of the
spectrometer. A peristaltic pump and a three-way valve were used to
circulate either the pure solvent or the surfactant solution through
the flow cell. Spectra were taken at a resolution of 4 cm$^{-1}$ for
8, 32 or 128 scans. A background spectrum was collected after the
cell was
filled with water, before the surfactant solution was pumped in
. Following
Harrick~\cite{Harrick60,Sperline87,Azzopardi94,Neivandt98,Singh01},
the amount of adsorbed surfactant can be quantified from the
absorbance of some vibration band of the molecule. In our case we
have followed the evolution of the CH$_2$ bands between 2800 and
3000 cm$^{-1}$. Absorbance of the vibration band $\nu_s$ (CH$_2$) at
2854 cm$^{-1}$ is used to determine the surface excess as a function
of time. This band has been chosen because it is less affected by
the baseline drift associated with the strong band of water in the
range 3200-3300 cm$^{-1}$. The surface excess $\Gamma$ is calculated
from~\cite{Harrick60,Azzopardi94}
\begin{equation}\label{Eq_Absorbance}
  A=k\epsilon\left[\frac{c_sd_p}{2}+\Gamma\right]
\end{equation}
\begin{equation}
  d_p=\frac{\lambda}{2\pi n_1\sqrt{\sin^2\theta-\left(\frac{n_2}{n_1}\right)^2}}
\end{equation}
\begin{equation}
  k=\frac{n_2{E_0}^2}{n_1\cos\theta}N
\end{equation}
where $d_p$ is the penetration length of the evanescent wave,
$\lambda$ the wavelength, $N$ the number of internal reflections,
$E_0$ the electric field amplitude, $n_1$ and $n_2$ the refractive
index of the germanium and the solution respectively, $\theta$ the
incident angle, $A$ and $\epsilon$ respectively the absorbance and
molecular extinction coefficient of the vibration band considered,
and $c_s$ the concentration of the absorbing species in solution.
Assumption is made that $c_s$ is not modified by adsorption. In
practice $N$ and $E_0$ cannot easily be determined so that $k$ is
determined from relation~(\ref{Eq_Absorbance}) by a calibration with
a non adsorbing compound of known extinction coefficient
(tert-butanol).
\subsubsection{AFM Surface forces measurement}
AFMs have been used for surface forces measurements in various
environments~\cite{Ducker92,Sounilhac99a,Sounilhac99b}.
Here the experiments were performed on a Nanoscope III (Digital Instrument)
with a silicon nitride tip using a liquid cell. Prior to the
experiment, the tip was cleaned by irradiation for 60 minutes in a
UV-ozone flow. A typical AFM experiment starts with a control of the
tip shape quality and the silica surface cleanliness by measuring
interaction forces between the AFM tip and the silica surface in
milli-Q water. The DOAIM solution is then introduced and the surface
forces profiles between tip and silica substrate are recorded every
3 minutes.
\subsubsection{Friction experiments}\label{Sec_Experimental_Friction}
\textcolor{black}{Friction experiments have been performed on two
reciprocating ball-on-plate tribometers: for low pressure friction
measurements, a home built millitribometer with a 50~mN load range
and a 0.02~mms$^{-1}$ maximum sliding velocity was used; for a
larger friction velocity range, a commercial (Plint T79) tribometer
with sliding velocity ranging from 0.01 to 10~mms$^{-1}$. However,
for this latter equipment, the normal load ranges between 0.1 and
20~N which results in larger mean pressures.} The plate is a silica,
2 mm thick substrate optically polished on both sides (GE quartz).
The ball is a fused silica sphere made from silica rods (GE quartz).
The end of the rod was melted with a blowtorch until a molten
droplet of glass formed with a radius of 2 to 4 mm. Both surfaces
were cleaned before use with a detergent-water-absolute ethanol
sequence for 15 minutes in an ultrasonic bath.

\begin{figure}\begin{center}
\includegraphics[width=12cm]{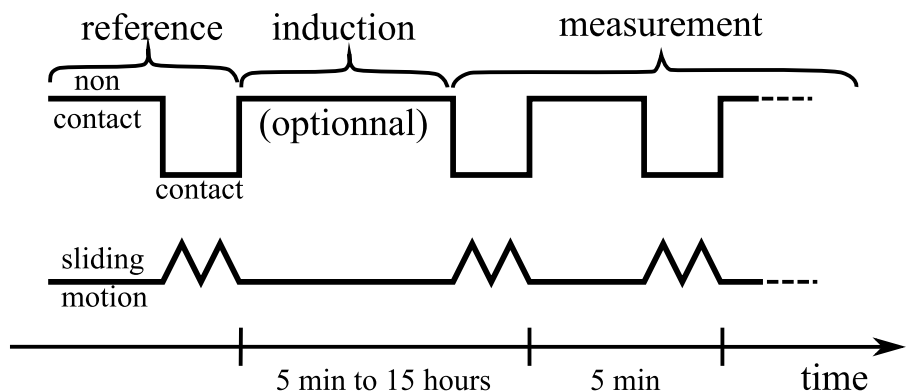}
\caption{\textcolor{black}{Experimental protocol for the measurement of the kinetics of the onset of lubrication. The friction measurement proceeds by short friction runs separated by 5 min intervals during which the surfaces are kept far apart but submerged in the solution, in order to avoid drying problems. The measurement period is optionally preceded by an induction period during which the surfaces are kept far apart in the solution, without contact or friction measurement. The reference are individual runs performed initially in water and immediately after introducing the solution.}}\label{Fig_Friction_Protocol}
\end{center}\end{figure}

\textcolor{black}{To emulate lubrication in the presence of the
sizing, the friction experiments were all conducted in the presence
of the aqueous surfactant dispersion, inside a liquid cell.} It is
also important to note that to minimize and control the impact of
shear, the typical friction experiments were not conducted as
continuous runs \textcolor{black}{as is usually done for such measurements}: on the contrary, unless otherwise stated, the
surfaces were brought into contact every 5 minutes for a series of
two cycles only, typically lasting a few seconds and were then
separated again (Fig.~\ref{Fig_Friction_Protocol}). The friction force was measured by averaging on the
second cycle. When separated, care was taken that the silica
surfaces remained immersed in the solution until the next
measurement. For each experiment, the friction coefficient was first
measured between surfaces immersed in pure water. The water was then
removed and replaced by the solution under study. The first point in
each friction graph is therefore the friction coefficient in pure
water.
%
%
\section{Results} 
\begin{figure}\begin{center}
\includegraphics[width=12cm]{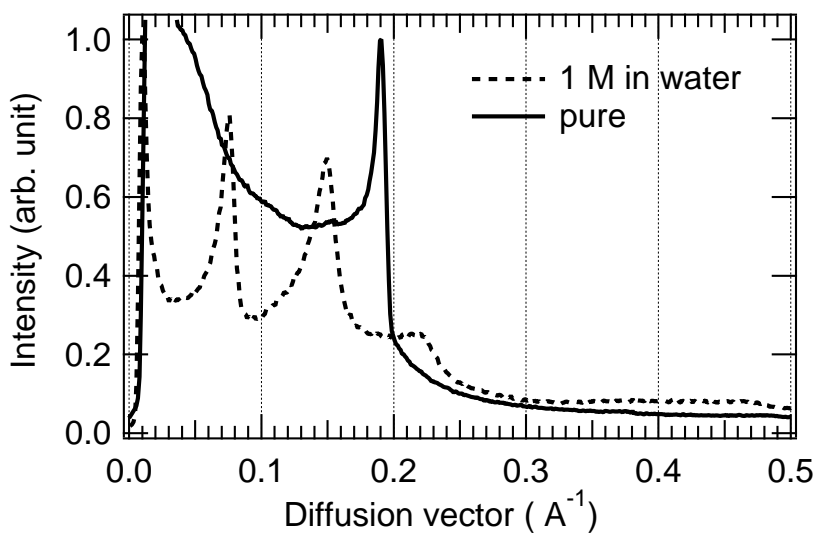}
\caption{\textcolor{black}{Xray diffraction spectra for pure and
diluted DOAIM. The fine diffraction peaks shifting to smaller wave
vector with dilution demonstrate the presence of a lamellar
phase.}}\label{Fig_RX}
\end{center}\end{figure}

\begin{figure}\begin{center}
\includegraphics[width=12cm]{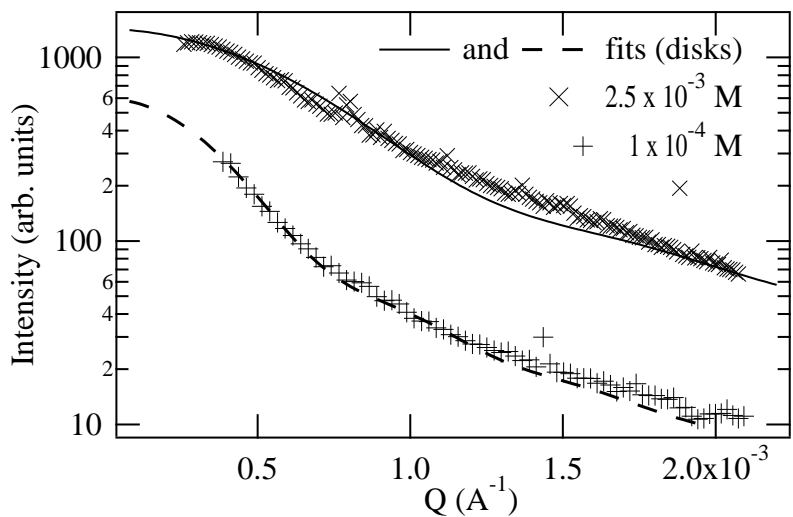}
\caption{\textcolor{black}{Static light scattering of DOAIM at low
concentrations with fits to diks shaped objects. The fits support
the expected extremely diluted bilayer structure at these low
concentrations.}}\label{Fig_SLS}
\end{center}\end{figure}

\textcolor{black}{
\subsection{Characterization of the solution and adsorption}
The pure surfactant system (after extraction of the isopropanol) is
optically birefringent. The SAXS diffractogram exhibits one single,
fine Bragg peak (Fig.~\ref{Fig_RX}) typical for an L$_\beta$ phase.
The repeat distance is 3.31~nm. After dilution in water (1~M), the
system exhibits shear induced birefringence which  persists over
days. In the SAXS diffractogram, a series of equally spaced peaks is
recorded (Fig.~\ref{Fig_RX}). These features are also typical for a
lamellar phase. The first order diffraction peak has moved to
smaller wave vector and the repeat distance has increased to
7.85~nm, which is fully consistent with the 7.65~nm value expected
for dilution of the lamellar phase to 1~M. Upon further dilution the
Xray signal and the optical birefringence is lost. Around
1$\times$10$^{-3}$~M, well above the CMC, the bilayer conformation is also
evidenced optically by the presence of multilamellar vesicles. At
lower concentrations, SLS experiments were carried out. The
scattered intensity recorded for 2.5$\times$10$^{-3}$ and 1.0$\times$10$^{-4}$~M
are displayed on Fig.~\ref{Fig_SLS}. Beyond the quadratic behaviour
for small diffusion wave vectors, the static correlation function
exhibits a moderate decay. The full shape of the correlation
function is consistent with extended disks~\cite{Zhmud05} \textcolor{black}{as expected for large dilutions
where the correlation between lamellae is lost}. The
measured correlation length are 310 and 550~nm for 2.5$\times$10$^{-3}$ and
1.0$\times$10$^{-4}$~M, showing that the materials behaves as sheets at
lengthscales smaller than the correlation length. In conclusion, the
surfactant solution exhibits a lamellar phase resulting from the
bilayer association of the individual surfactant molecules and the
bilayer structure is preserved upon dilution.}

\begin{figure}\begin{center}
\includegraphics[width=12cm]{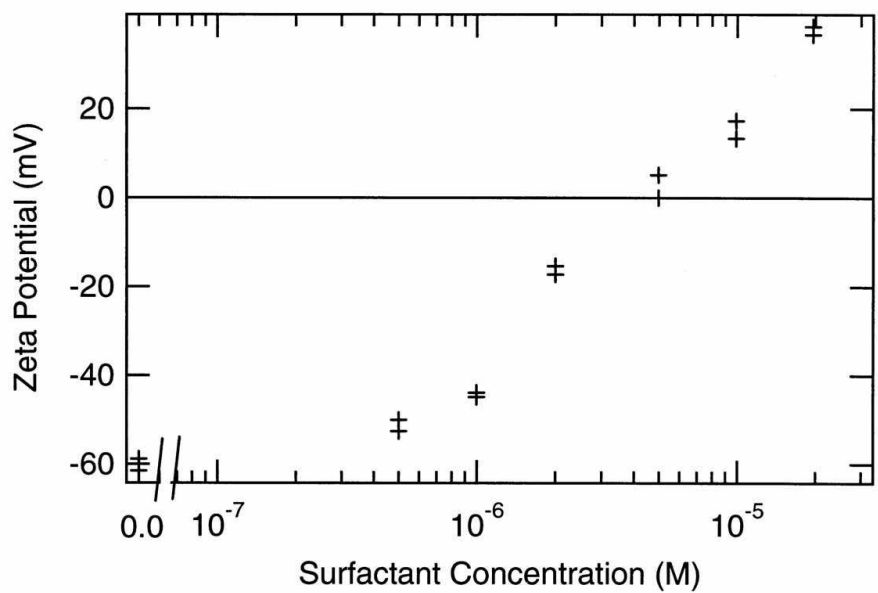}
\caption{\textcolor{black}{Zeta potential as a function of surfactant
concentration. The charge reversal is typical for the build-up of a
surfactant bilayer at the surface.}}\label{Fig_Zeta}
\end{center}\end{figure}

\textcolor{black}{From the surface tension as a function of
concentration, we determined a critical micelle concentration
CMC=9$\times$10$^{-6}$~M and an area per head $A_H$=0.71~nm$^2$. The
results of the zeta potential measurements at natural pH are
displayed on Fig.~\ref{Fig_Zeta}. The zeta potential of the bare
silica spheres is found at the expected -60 mV value. Upon
adsorption, the surface charge decreases and is finally reversed at
the point of zero charge PZC=5$\times$10$^{-6}$~M well below the CMC. This
charge reversal behavior is characteristic for the adsorption of a
bilayer at the surface.}

%
\subsection{Friction -- Time effect}
A first series of friction experiments were carried out at low
contact pressures (concentration C=10$^{-3}$~M, natural pH, sliding
velocity v=0.014~mms$^{-1}$, mean contact pressure P$_m$=90~MPa,
Figure~\ref{Fig_Friction_Time}). The typical friction coefficient of
silica surfaces immersed in pure water is 0.6$\pm$0.1 with a
variability due to surface preparation. Typical friction
coefficients after five minutes of immersion in the DOAIM solution
(C=10$^{-3}$~M) are down to 0.50 which indicates negligible (though
measurable) lubrication. On the other hand, if the surfaces are
first immersed 15 hours in a solution of DOAIM (10$^{-3}$~M) before
the friction experiment starts (procedure C) then the measured
friction coefficient is lower than 0.1, and sometimes reaches 0.03,
revealing fully lubricated surfaces. Similar results are recorded
for C=5$\times$10$^{-4}$~M. The long equilibration time in procedure C is
typical for surfactant lubrication
experiments~\cite{Helm92,Drummond03}. A friction coefficient of
about 0.1 or lower is usual for surfactant lubrication, especially
for double-chained surfactants with long
chains~\cite{Grant02,Yamada98,Briscoe06}. However, when the same
friction experiment starts immediately after immersion in the
surfactant solution (procedure A), the friction coefficient
decreases to reach the low friction value (about 0.05) after only
{\em ca} 2 hours. The transition is not linear with time. With
repeated friction tests carried out every 5 min we observe an
initial plateau at the high friction value around 0.5-0.6, which
typically lasts 1 hour before the friction coefficient starts to
decrease by one order of magnitude down to values around 0.05. These
results exemplify the fact that lubrication with double-chained
surfactants is not instantaneous but starts after an induction
period.
\subsection{AFM surface forces measurements}
\begin{figure}\begin{center}
\includegraphics[width=12cm]{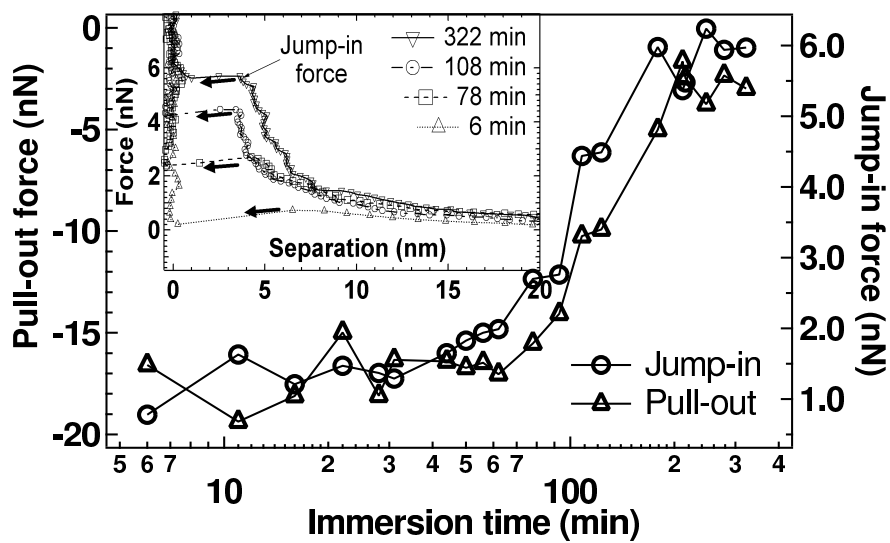}
\caption{Time evolution of the jump-in and of the pull-out forces
between a silicon nitride tip and a silica surface
(C=1$\times$10$^{-3}$~M). A few typical force vs distance curves are shown
as inset.}\label{Fig_AFM_Forces}
\end{center}\end{figure}

In the AFM force measurements, the force vs distance curve
(Figure~\ref{Fig_AFM_Forces}, inset) first displays a repulsive long
range interaction, followed by a steeper repulsive interaction
starting around 10 nm. \textcolor{black}{The former has not been
quantified due to the low signal to noise ratio but the results are
consistent with the electrostatic double layer interaction
demonstrated by the zeta potential measurements
(Fig.~\ref{Fig_Zeta}).} The latter is due to the mechanical
compression of the bilayer. Then, for distances close to 3-4 nm, the
AFM tip jumps into contact. This jump-in distance of 3-4 nm is close
to the thickness of a DOAIM bilayer. Such a behaviour is well-known
in the
literature~\cite{Subramanian01,Adler00,Rabinovich04,Rabinovich06}.
The jump-in force, defined as the repulsive force at jump-in, is an
estimate of the mechanical resistance of the bilayers. Pulling the
tip back induces the rupture of the tip-surface contact: a negative
force is registered which signals adhesion (not shown on the inset).
The amplitude of this pull-out force is a measure of the tip-surface
adhesion energy~\cite{Israelachvili,Barthel08}. The jump-in force in
a solution of DOAIM (1$\times$10$^{-3}$~M) has been measured as a function
of time, as well as the pull-out force
(Figure~\ref{Fig_AFM_Forces}). We note that the jump-in and the
pull-out forces measured here are tightly correlated, as already
reported in the literature~\cite{Helm92}. The jump-in force (counted
positive) increases as the magnitude of the pull-out force (counted
negative) decreases. They obey a time evolution similar to the
friction coefficient: in procedure A, it stays constant for about
one hour before the decrease to low friction; similarly jump-in and
pull-out forces exhibit an initial plateau before a transition
around 60 min to equilibrium values with large jump-in force and
negligible adhesion. \textcolor{black}{More complete results on the
adsorption isotherm measured by \textcolor{black}{ellipsometry} and the mechanical response
of the surfactant bilayer at equilibrium obtained with a Surface
Forces Apparatus will be published separately.}
\subsection{Adsorption kinetics of DOAIM}
\begin{figure}\begin{center}
\includegraphics[width=12cm]{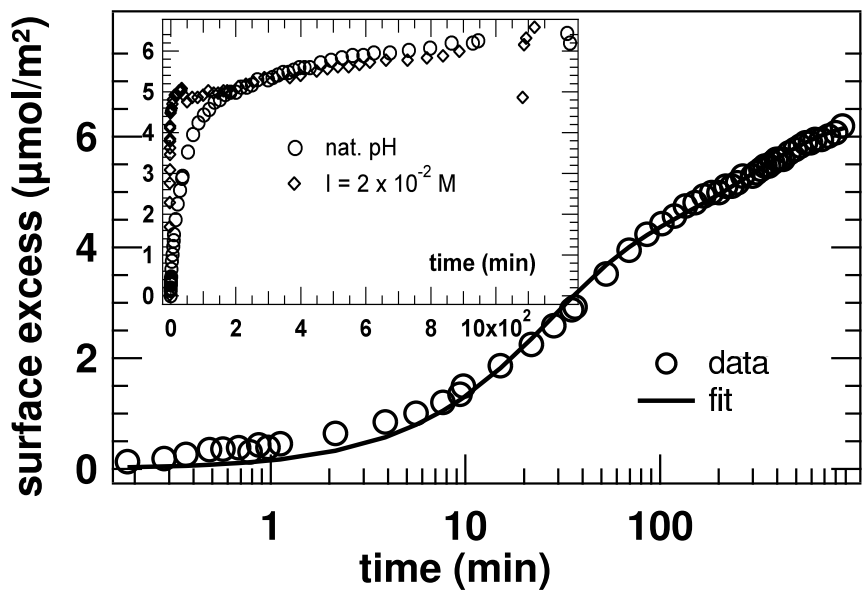}
\caption{Adsorption kinetics for DOAIM (5$\times$10~$^{-4}$~M) in pure
water and fit to a double exponential function ($\tau_1$= 25 min,
$\tau_2$=205 min). The inset compares the same data with the much
faster kinetics at high ionic strength (2$\times$10$^{-2}$~M) on a linear
time scale.}\label{Fig_IR_kinetics}
\end{center}\end{figure}

In this context, it is interesting to correlate the time evolution
of the contact properties with the amount of surfactant adsorbed on
the surface (surface excess). A measurement of the adsorption
kinetics (5$\times$10$^{-4}$~M, FTIR-ATR) is shown on
Figure~\ref{Fig_IR_kinetics}. Two different regimes are observed and
the data is reasonably well fitted by a double exponential function
with a fast time constant $\tau_1$=25 min and a slow time constant
$\tau_2$=205 min, leading to a pseudo-plateau. At pseudo-saturation,
the adsorbed amount is 6.2 $\mu$molm$^{-2}$. Rinsing with
recirculating water leads to little desorption, down to
5.5~$\mu$molm$^{-2}$. The area per molecule determined from surface
tension measurements is 0.79~nm$^2$. From this value we conclude
that at saturation, a full bilayer is formed at the silica surface.
A comparison of Figures~\ref{Fig_Friction_Time}
and~\ref{Fig_AFM_Forces} with Figure~\ref{Fig_IR_kinetics}
demonstrates that during the induction period, when the friction
coefficient is high and constant, the adsorption of the surfactant
is fast. When the transition to the lubricated state occurs
(procedure A), we can estimate that the surface excess is already
roughly as large as half a bilayer. From this observation we
conclude that there is no simple proportionality relation between
the adsorbed amount and the friction coefficient in this regime but
that a more complex mechanism is called for to explain the onset of
lubrication. In order to gain a clearer view of this mechanism, we
have performed a series of experiments to probe the impact of
kinetic parameters on the lubrication of the surfaces.
%
\subsection{Impact of shear on the onset of lubrication}\label{Sec_Shear}
We measured the friction when the experiment starts only three hours
after immersion in the solution (procedure B,
Figure~\ref{Fig_Friction_Time}). In such conditions, the adsorption
is almost complete (the surface excess amounts to 85\% of the
maximum, Figure~\ref{Fig_IR_kinetics}) and a lubricated surface is
obtained following procedure A. If lubrication were only controlled
by the adsorption of the surfactant, then a low friction coefficient
would be expected, as with procedure C. In contrast, a trend similar
to procedure A is observed: after 3 hours of induction, we measure
an initial value of the friction coefficient of approximately 0.5.
Transition towards a low friction coefficient is observed around 1.5
hours after the friction experiment has started, and a low value
(0.07) is reached about 2 hours after the beginning of the friction
experiment, that is a total of about 5 hours after immersion. This
result attests to the fact that shear accelerates the onset of
lubrication:
low friction is obtained after 2 hours in procedure A, in which
friction is probed every 5 minutes, but only after 5 hours in
procedure B, where the system is completely at rest for the first 3
hours.
\subsection{Ionic strength effect}
\begin{figure}\begin{center}
\includegraphics[width=12cm]{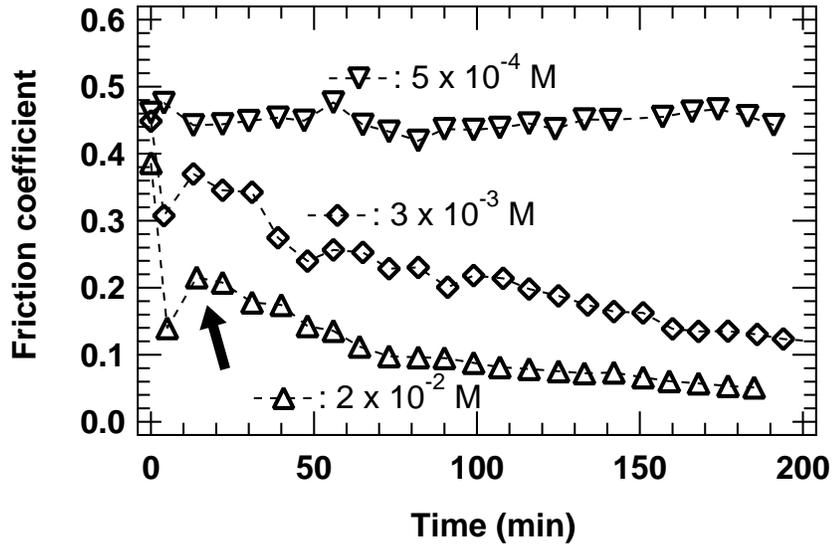}
\caption{Evolution of the friction coefficient versus immersion time
(C=5$\times$10$^{-4}$~M, natural pH) at different ionic strengths. The
arrow points to the friction spike observed after the initial
lubrication effect of the surfactant.}
\label{Fig_Friction_Ionic_Strength}
\end{center}\end{figure}

Ionic strength also impacts the adsorption process. In the presence
of salt (pH 4.5), the adsorption kinetics is much faster
(Figure~\ref{Fig_IR_kinetics}, inset) and for an ionic strength of
2$\times$10$^{-2}$~M, half-coverage of the surface is reached within
minutes. Figure~\ref{Fig_Friction_Ionic_Strength} shows the
evolution of the friction coefficient with time (DOAIM
5$\times$10$^{-4}$~M, P$_m$= 320-340~MPa, pH=4.5) for different salt
concentrations. The experiments follow procedure A where friction
starts immediately after immersion. We observe that the transition
towards low friction is considerably faster, a trend similar to the
adsorption kinetics. The initial friction plateau has now been
suppressed and the time to reach the lubricating state decreases
when the ionic strength increases. For the highest salt
concentration, the initial value of the friction coefficient after 5
minutes of immersion is already 3 times lower than in pure water.
Note that the friction spike to 0.2 which follows the first low
friction data point (Figure~\ref{Fig_Friction_Ionic_Strength}) is a
reproducible feature presumably connected to more dissipative
intermediate configurations towards a fully lubricated surface.
Similarly at high ionic strength the AFM force curves demonstrate an
almost instantaneous build-up of the repulsive force wall (not
shown).
\subsection{Sliding velocity}
\begin{figure}\begin{center}
\includegraphics[width=12cm]{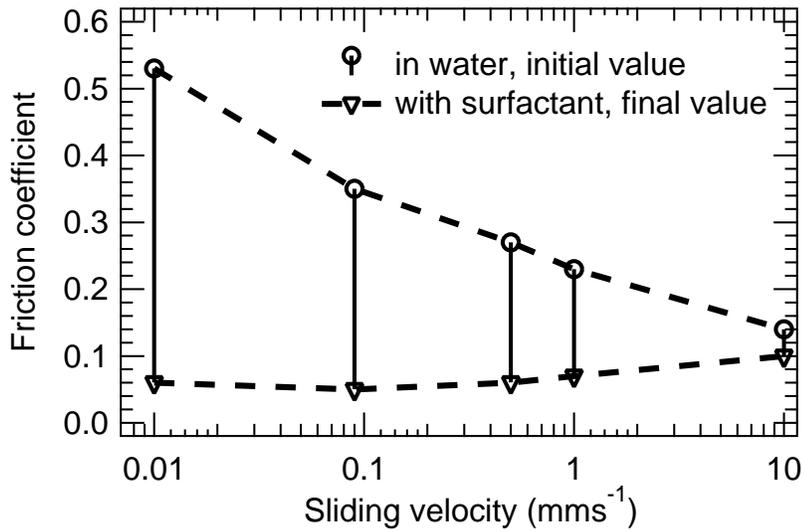}
\caption{Impact of shear velocity on the evolution of the friction
coefficient between silica surfaces (C=10$^{-3}$~M,
P$_m$=320-340~MPa, v=10~mms$^{-1}$).}\label{Fig_Friction_Velocity}
\end{center}\end{figure}

We have demonstrated in section~\ref{Sec_Shear} that shear
accelerates the transition to the lubricated state. A final set of
experiments aims at exploring the impact of {\it shear rate}.
Friction experiments at different sliding velocities were performed
(P$_m$=320-340~MPa) following procedure A. For the lower velocity
(0.01~mms$^{-1}$), the global evolution is similar to the evolution
recorded at lower mean pressure in contact (90~MPa): the high (0.53)
initial friction coefficient decreases with time to reach a stable
value of 0.06. \textcolor{black}{However, we can perform experiments
in a wider velocity range only at higher loads
(section~\ref{Sec_Experimental_Friction}). This is why} the
transition towards low friction coefficient is achieved after 5
hours instead of 2. For high velocities, the equilibrium
configuration is reached much faster, within 5-10~minutes.
Figure~\ref{Fig_Friction_Velocity} summarizes the main effects of
the sliding velocity on the initial and final values of the friction
coefficient. Note that the magnitude of the lubrication effect also
decreases, since the value of the friction coefficient of silica
surfaces in pure water decreases with the sliding
velocity~\cite{Toro04}.
\section{Discussion}
\subsection{The generic lubricating state}
\begin{figure}\begin{center}
\includegraphics[width=14cm]{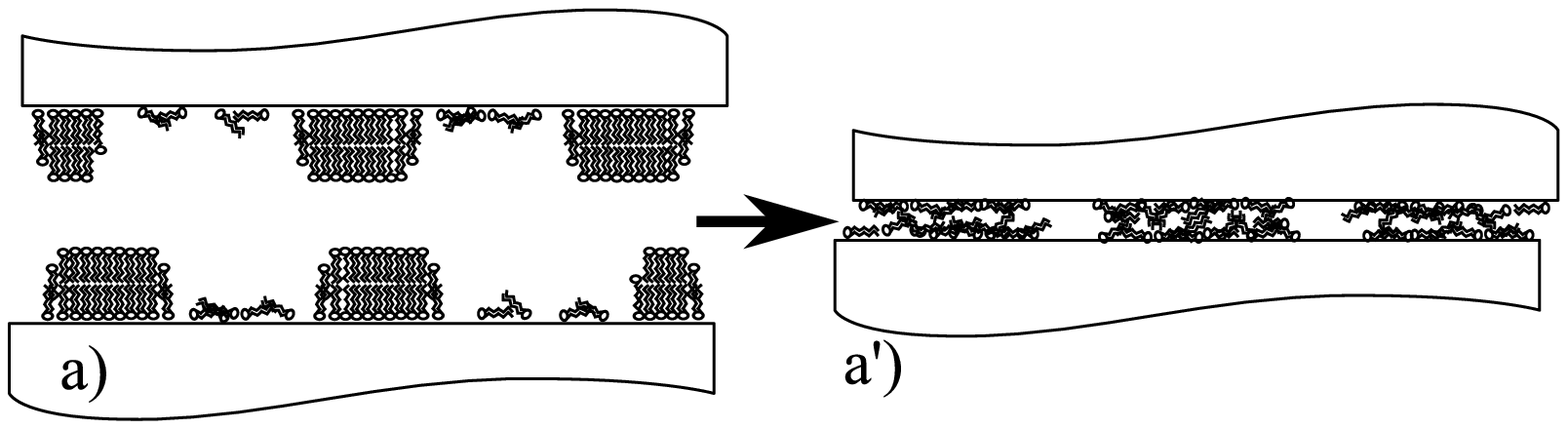}
\includegraphics[width=14cm]{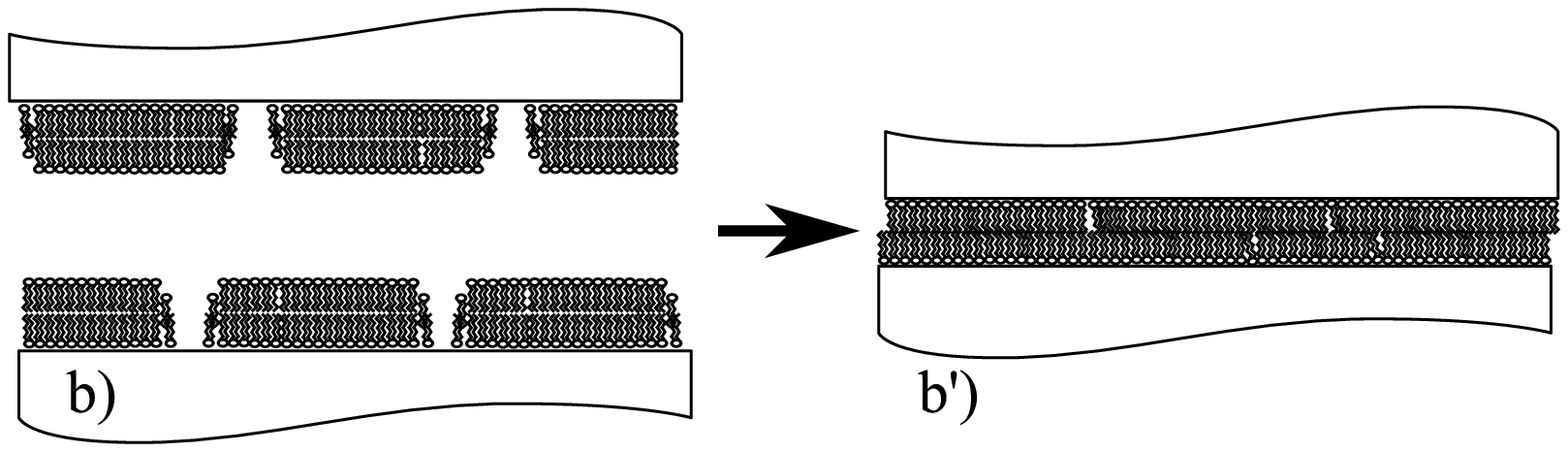}
\caption{Schematics of the surfactant instability for a highly
defective (a, a') and an almost defect-free (b, b')
bilayer.}\label{Fig_BiChain}
\end{center}\end{figure}

%
In the lubricated state, the friction coefficient is as low as 0.05,
as observed for instance in procedure C, after a 15 hour adsorption
period. For the types of loading used here, the mean normal
pressures in the contact are significantly larger than typical
hemifusion thresholds for double-chained C$_{18}$ surfactants
(several 10 MPa)~\cite{Helm89}. As a result, hemifusion of the
bilayers present on both surfaces occurs and this low value for the
friction coefficient results from the friction between two
hydrocarbon monolayers (Figure~\ref{Fig_BiChain} b'). In this
lubricated state, the average interfacial shear stress $\tau$ is of
the order of 20~MPa and is little affected by the sliding velocity.
Such a value is typical for monolayer-monolayer contacts in
air~\cite{Briscoe06} and is consistent with an approximate model
connecting friction and adhesion hysteresis
$\Delta\gamma$~\cite{Yamada98}. Indeed it has been proposed that
\begin{equation}\label{Eq_Friction}
\tau\simeq{\Delta\gamma}/{\delta}
\end{equation}
where $\delta$ is a molecular dimension. Reasonable values are
$\Delta\gamma\simeq 10$~mJm$^{-2}$ and $\delta\simeq
1$~nm~\cite{Briscoe06}, so that $\tau\simeq 10$~MPa. Friction
between the outermost surfaces of the two pristine bilayers would
lead to much lower friction coefficients: values one order of
magnitude lower, as low as 0.004, were reported for instance for
gemini surfactants~\cite{Drummond03}.
%
%
%
\subsection{Organisation during adsorption}
%
%
As amply demonstrated by our present results, this configuration is
not readily obtained upon adsorption from the solution. Indeed, the
equilibrium configuration in the bulk is usually different from the
equilibrium configuration of the surfactant aggregates adsorbed on a
surface \textcolor{black}{and the same surface excess may lead to very different surfactant conformations, with either dissipation and friction or lubrication}. Subtle effects control the \textcolor{black}{surfactant conformation} after adsorption
~\cite{Manne95,Wanless96,Lamont98,Atkin03,Paria04}.

This is especially true when the interaction is strong, which is the
case when surface and surfactant are oppositely charged: for
cationic surfactants, the electrostatic interaction with the
negatively charged silica surface results in fast initial adsorption
(Fig.~\ref{Fig_IR_kinetics}). After this first adsorption stage,
reorganization is required. For example, Chattoraj and
Biswas~\cite{Biswas98} observed two characteristics times for the
adsorption kinetics of short-chained cationic surfactants on silica
surfaces. The mechanism they propose is as follows: in the first
step, surfactant molecules from the bulk diffuse to the surface and
adsorb quickly with random orientation onto the silica surface; in
the second step, the crowded molecules tend to re-orient in a
regular fashion leading to the formation of adsorbed patches of
surface micellar aggregates. Such configuration changes will create
more vacant spaces for further adsorption of surfactant from the
bulk to the surface. Similarly, for adsorption of CTAB {\it above}
the cmc on mica surfaces, Chen {\it et al.}~\cite{Chen92} propose a
slightly different model where micelles adsorb directly on the
surfaces and subsequently reorganize. The idea is supported by the
fact that the same density of molecules is measured in the adsorbed
layer and in the micelles in solution.
%
%
\textcolor{black}{DOAIM, as many double-chained C$_{18}$ surfactant,
is dispersed as bilayers as further demonstrated by the present
optical, SAXS and SLS results.} An adsorption process similar to
lipid vesicle deposition must therefore be considered: the charged
vesicles present in the solution will adsorb quickly as patches of
bilayers and in an uncorrelated way~\cite{Richter03,Richter06}.
Rearrangement must proceed before a defect-free bilayer is
obtained~\cite{Neivandt98}. This scenario parallels the mechanism
proposed by Chen~\cite{Chen92} but here the rearrangement is
expected to be slower: the characteristic times for adsorption are
considerably larger than for short chain
surfactants~\cite{Biswas98,Subramanian01} since for long chain
surfactants, below the chain melting temperature, reorganization is
hampered by the slow dynamics ~\cite{Hayes98}.
%
%
\subsection{Bilayers, depletion and contact properties}
%
%
%
\textcolor{black}{The defective nature of the surfactant layer in the
initial stage of adsorption strongly impacts its mechanical
response. SFA experiments have shown that lipid bilayers exhibit a
smaller jump-in force and a larger pull-out force when depleted,
{\it i.e.} depletion facilitates hemifusion and increases
adhesion~\cite{Helm92}. Along the same line of thought, hemifusion
in the SFA has been shown to correlate with defect density
(monolayer or bilayer holes) as identified by AFM \cite{Benz04}. The
defect density was controlled by the deposition pressure in the
Langmuir-Blodgett trough. Similar studies have been reported for the
mechanical response of lipid bilayers measured by AFM as a function
of surface excess. AFM experiments have shown that
depletion~\cite{Grant02} and ionic strength~\cite{Oncins05} impact
friction. The results were somehow discussed in terms of packing
density. For lipid bilayers, an interesting suggestion is that the
reduced stability results from the increased hydrophobic
interactions between depleted bilayers~\cite{Helm92}, not from a
simple decrease in the density. Similarly, for shorter
single-chained surfactant it has also been observed either by SFA or
AFM force measurements that near the CMC, when the surfaces are
pushed to bilayer contact, the jump-in force increases with
surfactant concentration while the adhesion is maximum for monolayer
coverage~\cite{Singh01,Rutland94}. A connexion between micellization
energy and mechanical resistance at equilibrium has also been
established~\cite{Rabinovich06}.}

These observations all converge to demonstrate that an increase in
the packing density of molecules in the outer layer of the bilayer
leads to enhanced stability and reduction of adhesion. In the
present experiments the results demonstrate an increase of the
jump-in force and a reduction of adhesion as a function of time
(Fig.~\ref{Fig_AFM_Forces}). For DOAIM adsorbed at concentrations
significantly larger than the CMC, the initial conformation is
characteristic of frustrated aggregates adsorbed at the surface
(Fig.~\ref{Fig_BiChain}, a), which we loosely call defective
bilayer. Our results are in complete agreement with the picture of a
gradual healing of the initially defective bilayer.
\subsection{Friction and onset of lubrication}
%

%
Initially, before the defect-free bilayer is formed, a large
friction coefficient is recorded, around 0.5. Compared to the bare
silica-silica friction coefficient, this value demonstrate a very
moderate impact of the adsorbed surfactant. We suggest that this
sizeable interfacial shear results from the dissipation which
accompanies the deformation of the aggregates present at the
surface. These deformations may be transitions from bilayer to
tilted bilayer, aggregate ruptures, etc (Fig.~\ref{Fig_BiChain},
a'). Defective bilayers give rise not only to easier hemifusion and
enhanced adhesion, but also to friction because they allow for more
deformation at the molecular scale. \textcolor{black}{The results are
similar to the large friction recorded for lipid bilayers in AFM
experiments when a second mechanical transition threshold is
reached, well above hemifusion, and for which "direct surface
contact" is evoked~\cite{Oncins05,Grant02,Richter03}}.
%
%

\begin{figure}\begin{center}
\includegraphics[width=12cm]{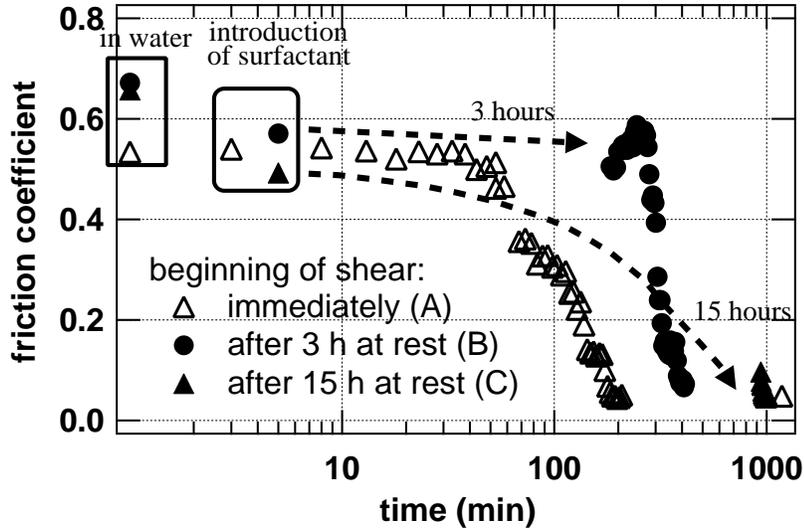}
\caption{Evolution of the friction coefficient vs immersion time
(C=10$^{-3}$~M, v=0.014~mms$^{-1}$, P$_m$=90~MPa). In all cases, the
friction coefficient is first measured in pure water, then
immediately after introduction of the surfactant. In procedure A,
the friction coefficient is measured every 5 min. Procedures B and C
are identical to procedure A, but the system is first left at rest
for respectively 3 and 15 hours before the friction coefficient
measurement starts. With procedures A and B, the friction is
initially high and stays constant for some induction period. Low
friction is observed immediately after the end of the 15 hour wait
period in procedure C.}\label{Fig_Friction_Time}
\end{center}\end{figure}

The decrease towards low friction is typically observed after 1 hour
(Fig~\ref{Fig_Friction_Time}). In parallel the surface forces
exhibit a decrease in the adhesion force and the repulsive jump-in
force becomes more pronounced (Fig.~\ref{Fig_AFM_Forces}). This
trend we connect with the organization at the surface which evolves
to a structure closer to a more ordered, stable, bilayer exposing
fewer hydrophobic moieties. The transition at the local scale from a
defective towards a stable bilayer has been completed. Indeed the
friction (Fig.~\ref{Fig_Friction_Time}) and adhesion
(Fig.~\ref{Fig_AFM_Forces}) drops recorded here are consistent with
Eq.~\ref{Eq_Friction}.
%
%
%
%

There remains to be explained why the transition is abrupt and does
not directly correlate with the surface excess. It is possible that
a critical flaw size exists below which the pressure-induced
transition to a tilted or disorganized layer is prevented. This
concept parallels the theory for bilayer
stability~\cite{Persson94,Barthel98b}. If the flaw density is large
enough, as occurs initially, the full surface induces dissipation
through aggregate edges. It is only when sufficient healing has
occurred and some defect-free patches have formed that the overall
friction coefficient decreases. In this scheme, the lubricated state
results from stabilisation of the surfactant layer through healing
of the larger defects.
%
%
%
%

The transition towards the lubricated state occurs faster in the
presence of salt, because ionic strength screens long range
electrostatic double layer interactions and facilitates
rearrangement.
Similarly, we have observed that the transition towards lubricating
state occurs earlier in time when the system is submitted to
friction immediately after immersion. Higher sliding velocity also
accelerates the transition. We conclude that shear and/or contact
due to the friction experiment itself favors the bilayer
organization of the surfactant between the two
surfaces~\cite{Rutland94}. \textcolor{black}{Indeed shear provides the
symetry breaking driving force which promotes
layering~\cite{Zipfel99,Li01} as well as the mechanical energy which
activates structural transitions~\cite{Akbulut05}}. It favours
surfactant accumulation and lamellar ordering turning the adsorbed
material into a fully formed bilayer~\cite{Boschkova02,Rutland94}.
\section{Conclusion}
%
%
The friction coefficient between millimetric silica surfaces was
measured during adsorption of a C$_{18}$ double-chained surfactant.
A transition from high to low friction is observed which parallels
the contact properties measured with the AFM. The results are not
directly correlated to the surface excess. They point to the role of
the organisation of the surfactant into a defect-free bilayer for
lubrication to be effective. Lubrication is obtained faster at
higher ionic strength and under shear because both facilitate the
bilayer organization. We have also study the impact of addition of
other surface active sizing components such as silanes on surfactant
lubrication. Strong effects have been evidenced due to interaction
and/or competitive adsorption, which have been published separately~\cite{Beauvais09}.

\section{Acknowledgements}
We thank M. Clerc-Imperor and R. Roquigny for the SAXS experiments.


%
\newpage


\end{document}